\documentclass[twocolumn,showpacs,preprintnumbers,amsmath,amssymb]{revtex4}
\usepackage{graphicx}% Include figure files
\usepackage{dcolumn}% Align table columns on decimal point
\usepackage{bm}% bold math

\begin{document}

% \preprint{APS/123-QED}

\title{Current induced entanglement of nuclear spins in quantum dots}

\author{Mikio Eto}
\affiliation{Faculty of Science and Technology, Keio University,
3-14-1 Hiyoshi, Kohoku-ku, Yokohama 223-8522, Japan}

\date{October 10, 2002}

\begin{abstract}
We propose an entanglement mechanism of nuclear spins in quantum dots
driven by the electric current. The current accompanied by the
spin flip in quantum dots gradually increases components
of larger total spin of nuclei. This entangled state drastically enhances
the spin relaxation rate of electrons, which can be detected by measuring
a leakage current in the spin-blockade region. This mechanism is not relevant
in optical experiments which examine the spin relaxation of single excitations.
\end{abstract}

\maketitle

The spin relaxation time of electrons in quantum dots is an
important issue. The time needs to be sufficiently long for the
implementation of quantum computing devices utilizing electron spins
in quantum dots \cite{Loss}.
The relaxation time has been found to be quite long by optical 
\cite{opt1,opt2} and transport experiments \cite{Fujisawa,Ono}.
In the former experiments, the transient measurements indicate a
quenching of the spin relaxation on the exciton lifetime scale in
self-assembled quantum dots. In the latter experiments,
the tunneling current is measured through single or double quantum dots
in ``spin blockade'' situations, where electrons cannot be
transported unless the spins are flipped in the dot.
The relaxation time is as long as $200$ $\mu$s from a spin-triplet
excited state to spin-singlet ground state in single quantum dots
\cite{Fujisawa}.
In weakly coupled double quantum dots, the current suppression has
been observed when two electrons occupy the lowest energy level in
each dot with parallel spins: an electron tunneling from one dot to
the other is forbidden by the Pauli exclusion principle \cite{Ono}.

In the present paper, we theoretically study a small leakage current
in spin-blockade regions. The current is accompanied by the spin flip
in the quantum dots.
As a spin-flip mechanism, we investigate the hyperfine interaction
between electrons and nuclear spins \cite{Sigurdur,Sigurdur2}.
Recent experimental results have implied
its important roles in the leakage current \cite{Ono2}.
We examine a formation of entangled states of nuclear spins driven
by the electron transport.
This mechanism is analogous to the current induced
dynamic nuclear polarization (DNP)
in quantum Hall systems \cite{DNP,Machida}. The DNP is created by
the electron scattering between spin-polarized regions, which leads
to a hysterisis in the longitudinal resistance \cite{Kronmuller,Smet}.
In our case, the entanglement of nuclear spins drastically enhances
the spin relaxation rate of electrons in quantum dots, which could
be observed experimentally. This mechanism is not relevant in the
optical experiments which examine the spin relaxation of single excitations.
Hence the role of the hyperfine interaction in the spin
relaxation could be quite different in transport and optical experiments.

{\it Model}: To consider the spin blockade in quantum dots,
we adopt a following model. A quantum dot is weakly coupled to
external leads L,R through tunnel barriers. The Coulomb blockade restricts
the number of electrons in the dot to be $N_{\rm el}$ or
$N_{\rm el}+1$. $N_{\rm el}$ electrons form a background of spin singlet.
From lead L on the source side, an extra electron tunnels into the dot 
and occupies a single level with envelope wavefunction
$\psi({\bf r})$. The spin of the electron is either $\uparrow$ ($S_z=1/2$)
or $\downarrow$ ($S_z=-1/2$) with equal probability
(we assume an easy-axis of electron spins in $z$ direction
for a while). The electron stays for a long time by the spin
blockade \cite{com0} because the spin relaxation time is much
longer than the tunneling time. After the spin flip in the dot,
the electron immediately tunnels out to lead R on the drain side,
and then the next electron is injected from lead L.

An electron occupying orbital $\psi({\bf r})$ interacts with $N$ nuclear spins,
${\bf I}_k$, by the hyperfine contact interaction.
$N \sim 10^5$ in GaAs quantum dots. We assume nuclear spins of
$1/2$ for simplicity. The Hamiltonian in the dot reads
\begin{equation}
H_{\rm hf} =  A \sum_{k=1}^N v_0 |\psi({\bf r}_k)|^2 {\bf S} \cdot {\bf I}_k
 \approx 2\alpha {\bf S} \cdot {\bf I},
\label{eq:Hamil}
\end{equation}
where $v_0$ is the volume of the crystal cell and
${\bf I}=\sum_{k=1}^N {\bf I}_k$ is the total spin of nuclei.
We have assumed that $\psi({\bf r})$ is independent of the nuclear site,
${\bf r}_k$.
This is a good approximation for a large part of nuclear spins in the quantum dot,
except in the vicinity of nodes of $\psi({\bf r})$,
since the distance between nuclei is much smaller than the size of the dot,
or an extension of $\psi({\bf r})$.

{\it Basic idea}:
The Hamiltonian (\ref{eq:Hamil}) indicates that $N$ nuclear spins interact
with a common ``field'' of electron spin although there is no direct
interaction between them (dipole-dipole interaction between nuclear spins
is weak and neglected). This field results in the entanglement among the
nuclear spins. To illustrate this, let us consider the simplest case of $N=2$
and begin with a polarized state of nuclear spins
$| I_{z}=1/2\rangle_1 | I_{z}=1/2 \rangle_2$. An electron with spin
$| \downarrow \rangle$ tunnels into the dot and is spin-flipped by the
hyperfine interaction. Then the state of nuclear spins becomes
$(| -1/2 \rangle_1 | 1/2 \rangle_2 + | 1/2 \rangle_1 | -1/2\rangle_2 )
/\sqrt{2} =| J=1,M=0 \rangle$, where $J$ and $M$ are the total spin and its
$z$ component, respectively. This is an entangled state.
After the electron tunnels out of the dot, the next electron is injected with
$| \downarrow \rangle$ (or $| \uparrow \rangle$). The spin-flip
probability of the second electron is proportional to
$|\langle 1, \mp 1; \uparrow (\downarrow) |H_{\rm hf}|
1,0; \downarrow (\uparrow) \rangle |^2 =2\alpha^2$.
This value is twice the probability in the case of non-entangled states
$|1/2\rangle_1 | -1/2 \rangle_2$ or $|-1/2\rangle_1 | 1/2 \rangle_2$.
In general, the capability of state $|J,M \rangle$ to
flip an electron spin is $\propto (J \pm M)(J \mp M +1)$.

{\it In presence of easy-axis}:
We formulate the entanglement of $N$ nuclear spins. In the initial state,
they are not entangled at all and oriented randomly
\begin{eqnarray}
\Psi^{(0)} & = & (c_1 |1/2 \rangle_1 +d_1 |-1/2 \rangle_1) 
\otimes \cdots \nonumber \\
& \otimes & (c_N |1/2 \rangle_N+d_N |-1/2 \rangle_N).
\label{eq:Psi00}
\end{eqnarray}
$\{ c_k, d_k \}$ are randomly distributed $(|c_k|^2+|d_k|^2=1)$.
By the transformation of the basis set from
$\{ |I_z \rangle_1 \otimes |I_z \rangle_2 \otimes \cdots \}$ to the
total spin of all the nuclei, $\{ | J,M \rangle \}$ ($J \le N/2$,
$-J \le M \le J$), Eq.\ (\ref{eq:Psi00}) is rewritten as
\begin{equation}
\Psi^{(0)}=\sum_{J,M,\lambda} C_{J,M,\lambda}^{(0)} | J,M,\lambda \rangle,
\label{eq:Psi0}
\end{equation}
where the coefficients $\{ C_{J,M,\lambda}^{(0)} \}$ are randomly distributed
($\sum_{J,M,\lambda} |C_{J,M,\lambda}^{(0)}|=1$). Index
$\lambda$ distinguishes states with the same $J$ and $M$.
The number of such states is given by
\begin{equation}
K(N,J)=(2J+1)\frac{N!}{(\frac{N+2J+2}{2})! (\frac{N-2J}{2})!}.
\end{equation}
After the injection of the first electron, say, $| \downarrow \rangle$,
the time evolution of the dot state by $H_{\rm hf}$ is
\begin{widetext}
\begin{eqnarray}
e^{-iH_{\rm hf}t/\hbar}\Psi^{(0)}\otimes | \downarrow \rangle
=\sum_{J,M,\lambda} C_{J,M,\lambda}^{(0)}
\frac{e^{-i\alpha Jt/\hbar}}{2J+1}\left[ J-M+1 + 
(J+M)e^{i\alpha (2J+1) t/\hbar} \right] | J,M,\lambda \rangle
\otimes | \downarrow \rangle
\nonumber \\
+\sum_{J,M,\lambda} C_{J,M,\lambda}^{(0)}
\sqrt{(J+M)(J-M+1)}
\frac{e^{-i\alpha Jt/\hbar}}{2J+1}
(1 - e^{i\alpha (2J+1) t/\hbar}) | J,M-1,\lambda \rangle
\otimes | \uparrow \rangle.
\label{eq:t-evolve}
\end{eqnarray}
\end{widetext}
The expansion to the lowest order in $\alpha t/\hbar$ yields
the spin-flip probability, $P^{(0)}=F^{(0)}(\alpha t/\hbar)^2$, with
\begin{equation}
F^{(0)}=\sum_{J,M,\lambda} |C_{J,M,\lambda}^{(0)}|^2(J \pm M)(J \mp M+1)
\end{equation}
(the lower sign indicates the electron spin flip from
$| \uparrow \rangle$ to $| \downarrow \rangle$).
When $N \gg 1$, $|C_{J,M,\lambda}^{(0)}|^2$ can be replaced by
$1/2^N$ (law of large number). Then
$F^{(0)}=(1/2^N)\sum_{J,M} K(N,J) (J \pm M)(J \mp M+1)=N/2$.
$P^{(0)}=(N/2)(\alpha t/\hbar)^2$ is identical to the spin-flip probability
which would be evaluated on the assumption that one of the nuclear spins is
flipped with the electron spin.

The electron tunnels off the dot immediately after the spin is flipped.
Then the wavefunction $\Psi^{(0)}$ shrinks to
\begin{eqnarray}
\Psi^{(1)} = \frac{1}{\sqrt{F^{(0)}}}\sum_{J,M,\lambda} C_{J,M,\lambda}^{(0)}
\sqrt{(J \pm M)(J \mp M+1)}
\nonumber \\
\times | J,M \mp 1,\lambda \rangle.
\end{eqnarray}
$\Psi^{(1)}$ is partly entangled since it contains more components of
larger $J$ and smaller $|M|$. The degree of the entanglement increases
each time an electron is injected, spin-flipped, and ejected out of the dot.
After $n$ events, the state of nuclear spins becomes
\begin{eqnarray}
\Psi^{(n)} & = & \sum_{J,M,\lambda} C_{J,M,\lambda}^{(n)}
                | J,M,\lambda \rangle, \\
C_{J,M \mp 1,\lambda}^{(n)} & = &
\sqrt{\frac{(J \pm M)(J \mp M+1)}{F^{(n-1)}}} C_{J,M,\lambda}^{(n-1)},
\label{eq:Psin}
\\
F^{(n)} & = &
\sum_{J,M,\lambda} |C_{J,M,\lambda}^{(n)}|^2 (J \pm M)(J \mp M+1).
\end{eqnarray}
$F^{(n)}$ is expressed as
\begin{eqnarray}
F^{(n)} & = & f^{(n)} / f^{(n-1)}, \nonumber \\
f^{(n)} & = & \frac{1}{2^N}\sum_{J,M} K(N,J) \left[ (J \pm M)(J \mp M+1) \right]^n.
\label{eq:Fn}
\end{eqnarray}
Figure 1 shows the distribution of (a) the total spin $J$ and 
(b) its $z$ component $M$. The spin-flip processes increase the weight of
larger $J$ and smaller $|M|$. Although $M$ changes less faster than $J$, this means
that the total nuclear spins tend to develop
in the plane perpendicular to the easy-axis of electron spins.

%%%%%%%%%%%%%%
\begin{figure}
\includegraphics[width=6cm]{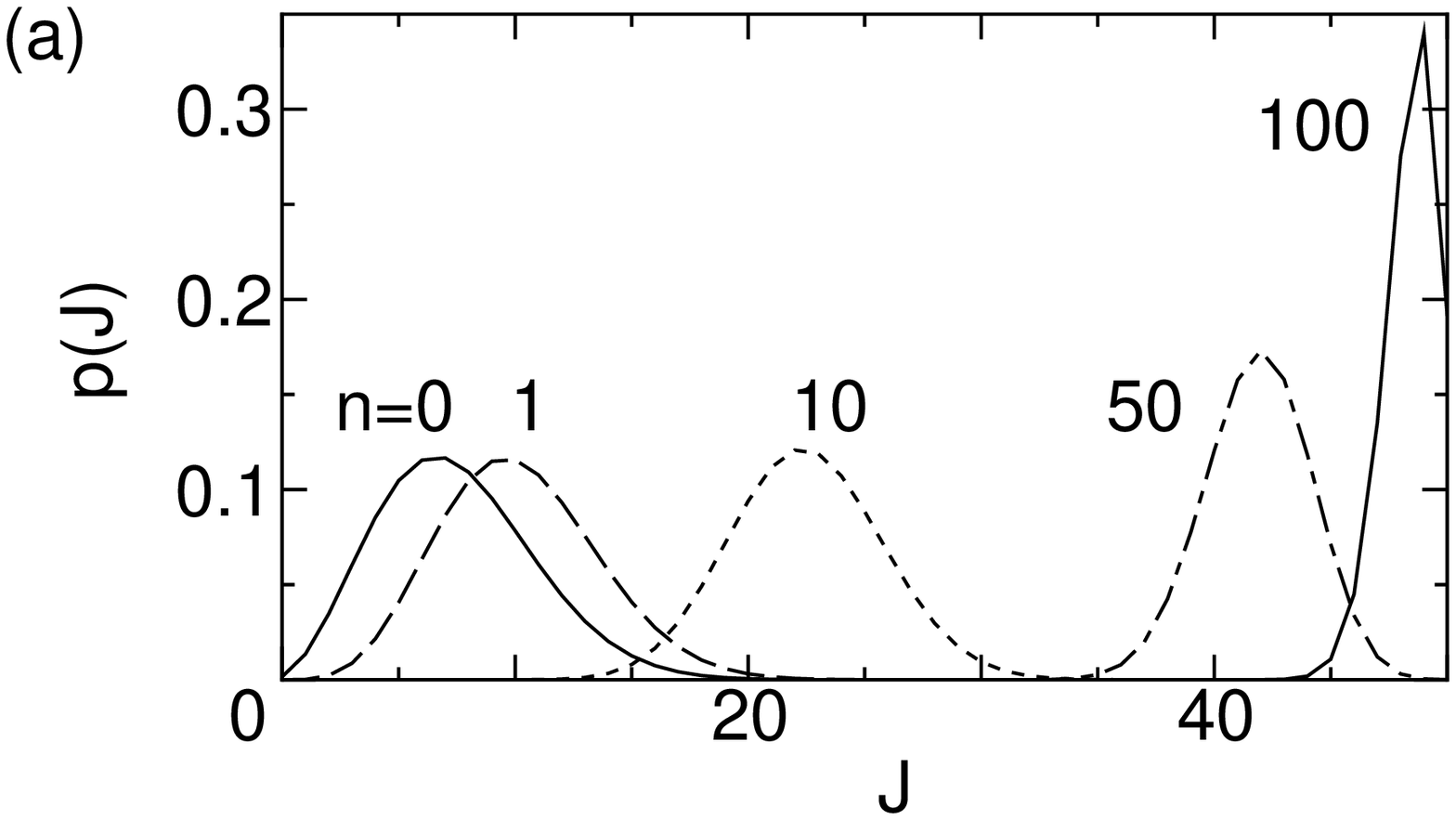} \\
\includegraphics[width=6cm]{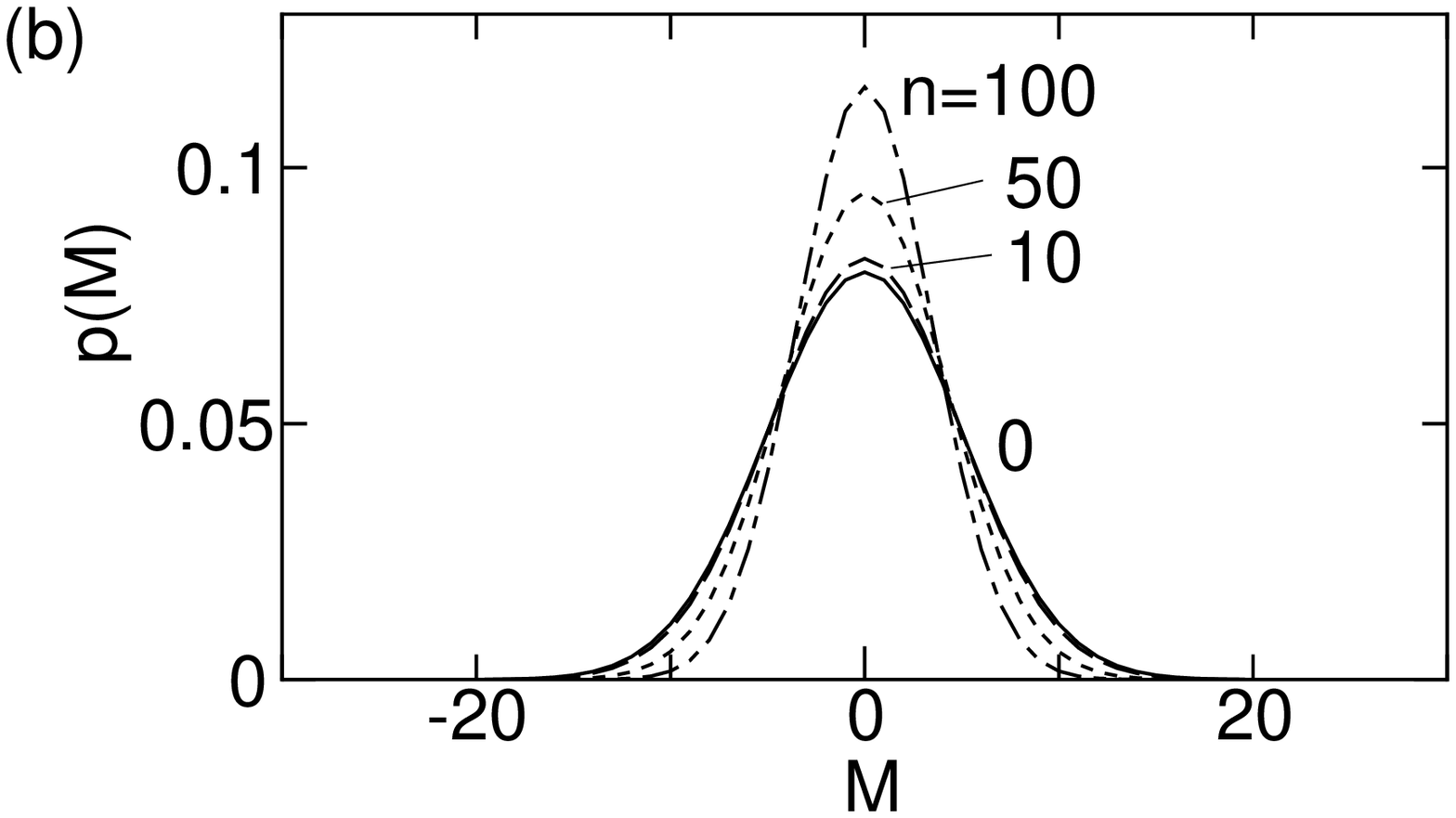}
\caption{The distribution of (a) the total spin $J$ and (b) its $z$
component, in the state of nuclear spins. An easy-axis of electron
spins is assumed.
(a) $p(J)=\sum_{M,\lambda} |C_{J,M,\lambda}^{(n)}|^2$, and
(b) $p(M)=\sum_{J,\lambda} |C_{J,M,\lambda}^{(n)}|^2$, where
$n$ is the number of transported electrons accompanied by the spin flip.
The number of nuclear spins is $N=100$. In Eq.\ (\ref{eq:Psin}),
we take the geometrical mean between upper and lower signs.}
\end{figure}
%%%%%%%%%%%%%%
This entangled state enhances the spin-flip probability.
For the $(n+1)$th electron,
the probability is given by $P^{(n)}=F^{(n)}(\alpha t/\hbar)^2$.
$P^{(n)}$ with fixed $t$ is shown by solid
line in Fig.\ 2(a). Using Eq.\ (\ref{eq:Fn}), we find that
$F^{(n)} \approx (N/2)n$ for $1 \ll n \ll N/2$, and
$F^{(n)} \approx (N/2)^2$ for $N/2 \ll n$. Therefore the probability
increases with $n$ linearly
($P^{(n)} \approx n P^{(0)}$) and finally saturates
($P^{(n)} \approx (N/2) P^{(0)}$).

{\it In absence of easy-axis}:
Until now, we have assumed the presence of easy-axis of electron spins.
In the absence of the axis, the spin of the $n$th electron is
oriented in an arbitrary direction, $\cos\theta^{(n)} | \uparrow \rangle + 
\sin\theta^{(n)}e^{i\varphi^{(n)}}| \downarrow \rangle$.
In this case, the previous formulation can be applied after the
rotation of $\{ C_{J,M,\lambda}^{(n)} \}$,
$C_{J,M,\lambda}^{(n) \prime}=\sum_{M'=-J}^{J}
U(J, \theta^{(n)},\varphi^{(n)})_{M,M'} C_{J,M',\lambda}^{(n)}$,
to align the $z$ axis parallel to the spin direction.
If the spin direction of incident electrons is completely
random, we can average over the $z$-component of the total spin.
Then the coefficients $\{ C_{J,M,\lambda}^{(n)} \}$ in Eq.\ (\ref{eq:Psin})
are replaced by the averaged values $\{ \bar{C}_{J,\lambda}^{(n)} \}$.
They develop by
\begin{equation}
\left| \bar{C}_{J,\lambda}^{(n)} \right|^2=
\frac{1}{\sqrt{G^{(n-1)}}} \frac{2}{3}J(J+1)
\left| \bar{C}_{J,\lambda}^{(n-1)} \right|^2,
\label{eq:Psi2n}
\end{equation}
where
\begin{eqnarray}
G^{(n)}  & = & g^{(n)} / g^{(n-1)}, \nonumber \\
g^{(n)} & = & \frac{1}{2^N}\sum_{J} K(N,J) (2J+1) \left[ \frac{2}{3}J(J+1) \right]^n.
\label{eq:Gn}
\end{eqnarray}
As in the previous case, components of larger $J$ increase with $n$,
which enlarge the spin-flip probability by the factor of $2J(J+1)/3$.
The probability is given by $P{(n)}=G^{(n)}(\alpha t/\hbar)^2$, which
is shown by broken line in Fig.\ 2(a). This behaves as
$P^{(n)} \approx (2n/3) P^{(0)}$ for $1 \ll n \ll N/2$, and
$P^{(n)} \approx (N/3) P^{(0)}$ for $N/2 \ll n$, with fixed $t$.

%%%%%%%%%%%%%%
\begin{figure}
\includegraphics[width=6cm]{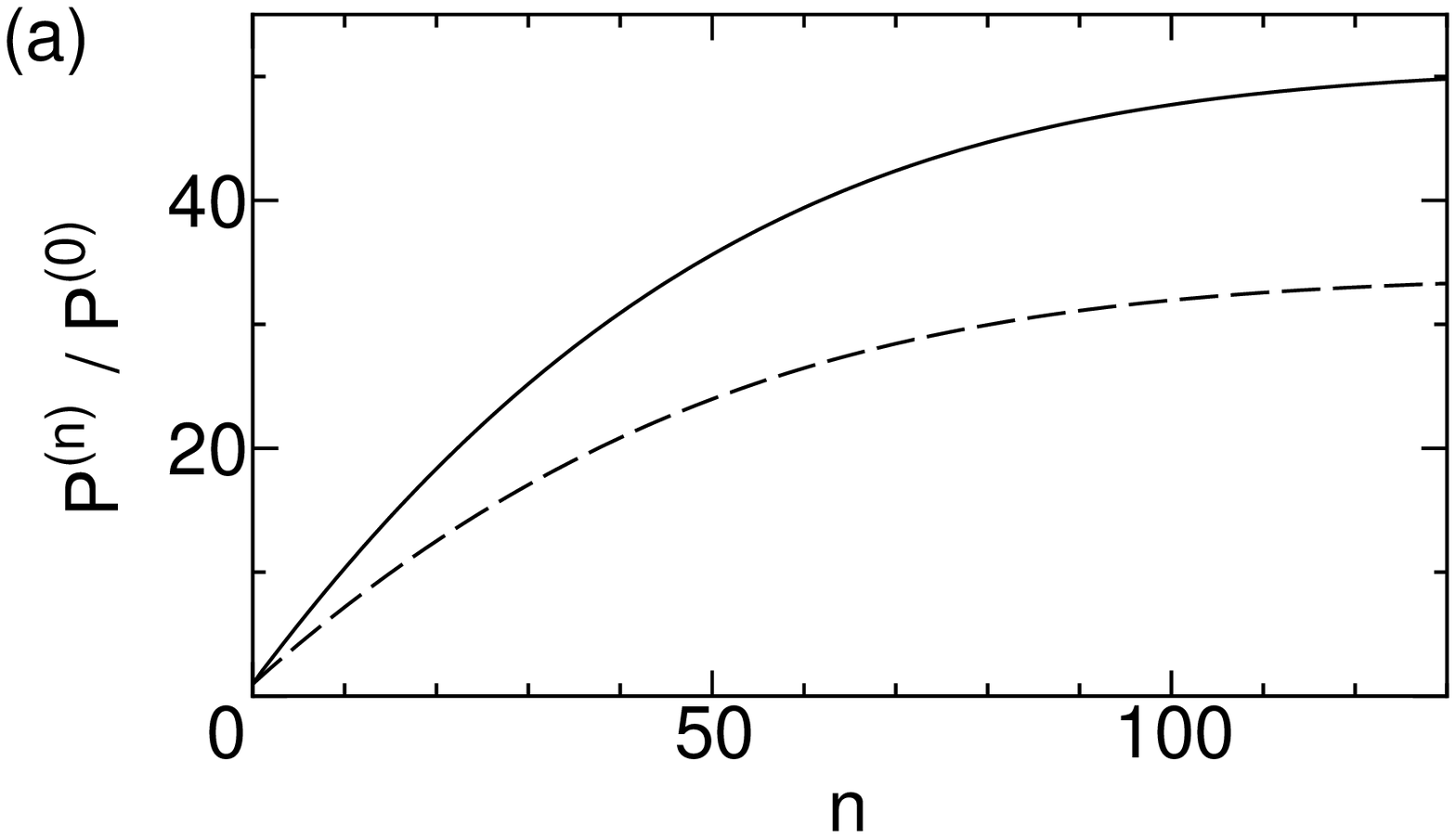} \\
\includegraphics[width=6cm]{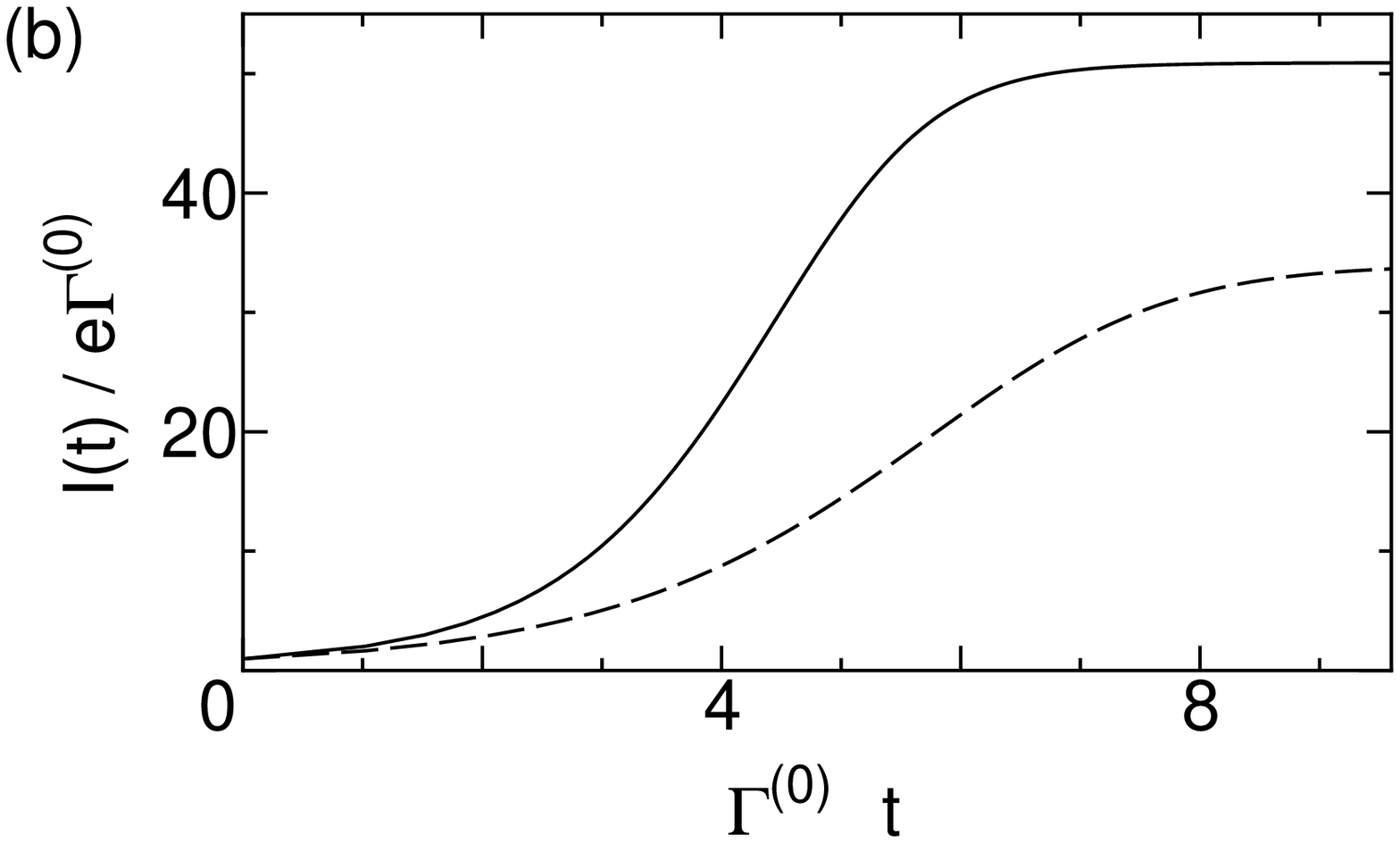}
\caption{(a) The spin-flip probability $P^{(n)}$ with fixed $t$,
as a function of $n$ (number of transported electrons accompanied by
the spin flip). (b) The spin-flip current $I(t)$
as a function of time $t$. The cases in the presence and absence of
easy-axis of electron spins are indicated by solid and broken lines,
respectively.
The number of nuclear spins is $N=100$. In the presence of easy-axis,
we take the geometrical mean between upper and lower signs in
Eq.\ (\ref{eq:Fn}).}
\end{figure}
%%%%%%%%%%%%%%

{\it Leakage current}:
The enhancement of the spin-flip probability is reflected by the
leakage current. Instead of calculating the current in the previous situations,
now we consider the case in the presence of magnetic field: Zeeman energy for
electrons is much larger than the hyperfine interaction, whereas that
for nuclear spins is negligible. Then $H_{\rm hf}$ can be treated as a
perturbation. The magnetic field makes an easy-axis of electron spins.
We also take into account the electron-phonon interaction
for the energy conservation at the spin flip \cite{Sigurdur2}.
The Hamiltonian in the dot reads
$H=H_{\rm el}+H_{\rm ph}+H_{\rm hf}+H_{\rm el-ph}$, where
$H_{\rm el}=-E_{\rm Z}(a^{\dagger}_{\uparrow}a_{\uparrow}-
a^{\dagger}_{\downarrow}a_{\downarrow})$,
$H_{\rm ph}=\sum_{\bf q}\hbar\omega_{\bf q} b^{\dagger}_{\bf q}b_{\bf q}$,
and
$H_{\rm el-ph}=\sum_{\sigma}\sum_{\bf q} \beta_{\bf q}
a^{\dagger}_{\sigma}a_{\sigma} (b^{\dagger}_{\bf q}+b_{\bf q})$, with
$a^{\dagger}_{\sigma}$, $a_{\sigma}$ ($b^{\dagger}_{\bf q}$, $b_{\bf q}$)
being creation and annihilation operators for an electron in the dot
(phonon). The spin-flip processes in the lowest order
in $H_{\rm hf}+H_{\rm el-ph}$
(first orders in $H_{\rm hf}$ and in $H_{\rm el-ph}$) change the
nuclear spin state in the same way as in Eq.\ (\ref{eq:Psin}) \cite{com1}.

The spin-flip rate for the $(n+1)$th electron is
$\Gamma^{(n)}\propto F^{(n)}$. The time interval between $n$th and
$(n+1)$th electron transports is $\Delta t^{(n)} \approx 1/\Gamma^{(n)}$.
Thus the current is given by
$I(t)=e\Gamma^{(n)}$ at $t=\sum_{j=0}^{n} \Delta t^{(j)}$. This current
is shown in Fig.\ 2(b), as a function of $t$ (solid line).
The approximate form of $F^{(n)}$ yields an asymptotic form of the current
\begin{equation}
\left\{
\begin{array}{llll}
I(t) & \approx & e\Gamma^{(0)} e^{\Gamma^{(0)} t}  & \hspace{.5cm}
(t \ll t_{\rm sat}) \\
I(t) & \approx & e(N/2)\Gamma^{(0)} & \hspace{.5cm} (t_{\rm sat} \ll t).
\end{array}
\right.
\end{equation}
The current grows with time drastically and finally saturates.
The saturation time is given by
$t_{\rm sat}=\ln(N/2)/\Gamma^{(0)}$, where $\Gamma^{(0)}$ is the spin-flip
rate of an electron with the non-entangled state of nuclear spins.
The increase in the current continues until the entangled state is broken
by dephasing effects.

Even in the absence of magnetic field, the electron energies for
$| \uparrow \rangle$ and $| \downarrow \rangle$ could be split
in $H_{\rm el}$, $e.g$.\ by spin-singlet or -triplet formation
with an electron trapped in the other dot, in double dot systems.
In this case, the easy-axis of electrons would be absent
(if the crystal field is weak enough).
The perturbation with respect to $H_{\rm hf}+H_{\rm el-ph}$
changes the nuclear spin state following Eq.\ (\ref{eq:Psi2n}).
The spin-flip rate is $\Gamma^{(n)}\propto G^{(n)}$.
The leakage current $I(t)$ is represented by
broken line in Fig.\ 2(b). It is written approximately as
\begin{equation}
\left\{
\begin{array}{llll}
I(t) & \approx & e(2\Gamma^{(0)}/3) e^{2\Gamma^{(0)} t/3}  & \hspace{.5cm}
(t \ll t_{\rm sat}) \\
I(t) & \approx & e(N/3)\Gamma^{(0)} & \hspace{.5cm} (t_{\rm sat} \ll t),
\end{array}
\right.
\end{equation}
where $t_{\rm sat}=3\ln(N/2)/(2\Gamma^{(0)})$.
The enhancement of the current is less prominent than
in the presence of easy-axis.

{\it Discussion}:
We have proposed the current induced entanglement of nuclear spins in
quantum dots. The current accompanied by the spin flip in the dots
gradually increases components of larger total spin of nuclei,
which significantly enhances the spin relaxation of electrons.
As a result, the leakage current in the spin-blockade region grows
drastically with time and finally saturates.
This mechanism is not relevant in the optical experiments which examine
the spin relaxation of single excitations.

The saturation time of the leakage current is of the same order
as $1/\Gamma^{(0)}$, spin relaxation time with non-entangled state
of nuclear spins, which is $\sim 100$ ns or much larger \cite{Fujisawa,Ono}.
The current is enhanced during the dephasing time $T_2$.
When the entangled state is broken, the current is suppressed.
Possibly this would result in a current fluctuation with time which
has been observed recently \cite{Ono2}. The characteristic time of the
fluctuation is $T_2$.

A possible origin of the dephasing is the dipole-dipole interaction
between nuclear spins. Note that, among $N$ nuclear spins participating in
the formation of total spin, $|J,M \rangle$, this interaction conserves
the total spin, and hence does not destroy the entangled state.
The interaction between one of the $N$ spins and a spin
surrounding the dot results in the dephasing.
Estimating $T_2$ in our case and analyzing evolution of nuclear
spins after the dephasing are beyond the present paper.

We discuss the validity of our simple models.
We have disregarded the spatial variation of the envelope
wavefunction of electrons in the quantum dot.
Although the entanglement of nuclear spins is generally seen
in the presence of the variation
(see Eq.\ (4) in Ref.\ \cite{Khaetskii2}), the enhancement of the
spin relaxation is more effective with larger total spin $J$.
Hence the capability to flip an electron spin is determined by
the effective number $N_{\rm eff}$ of nuclear spins which feel the
same electron field, $\psi({\bf r})=$ const.,
since $J_{\rm max}=N_{\rm eff}/2$. The generalized evolution of
nuclear spins is an interesting problem.

If the contribution from higher energy levels of electrons
in quantum dots is not negligible, through electron-phonon interaction
\cite{Sigurdur,Sigurdur2} or spin-orbit (SO) interaction \cite{Khaetskii},
it should be taken into account carefully.
Particularly, the coexistence of hyperfine and SO interactions would
complicate the evolution of the nuclear spin state
because the terms of $| \downarrow \rangle$ and $| \uparrow \rangle$
are mixed on the right side of Eq.\ (\ref{eq:t-evolve}).
We have also disregarded higher-order tunneling processes \cite{Averin},
which could play a role in the spin relaxation of electrons
\cite{Fujisawa}. However, the processes do not influence the entanglement of
nuclear spins discussed here.

Finally, we comment an analogy between our mechanism and the Dicke effect of
superradiance \cite{Dicke,Brandes}. The spontaneous emission of
photons is enhanced from $N$ atoms with two levels (pseudo-spin $S_z=\pm 1/2$)
if all of them are excited initially. This is due to the formation of
pseudo-spin state $|J,M \rangle$ with $J=N/2$. (Starting from  $|J,J \rangle$,
the state of $N$ atoms changes like a cascade, $|J,J \rangle$,
$|J,J-1 \rangle, \cdots$, emitting photons.)
A similar effect has been proposed for the emission of phonons from $N$
equivalent quantum dots \cite{Brandes,Brandes2}.
The atoms (quantum dots) correspond to the
nuclear spins in our model, while the emission of photons (phonons) to
the spin flip of electrons. A main difference is the initialization.
$N$ excited states have to be prepared by the pumping in the Dicke effect,
whereas the initialization is not necessary in our mechanism.

The author gratefully acknowledges discussions with
K.\ Kawamura, R.\ Fukuda, S.\ Komiyama, T.\ Inoshita, A.\ Shimizu, and
K.\ Ono.
Numerical calculations for Figs.\ 1 and 2 were performed by M.\ Murata.


\begin{references}
\bibitem{Loss}
D.\ Loss and D.\ P.\ DiVincenzo, Phys.\ Rev.\ A {\bf 57}, 120 (1998).
\bibitem{opt1}
M.\ Paillard {\it et al}., Phys.\ Rev.\ Lett.\ {\bf 86}, 1634 (2001).
\bibitem{opt2}
A.\ S.\ Lenihan {\it et al}., Phys.\ Rev.\ Lett.\ {\bf 88}, 223601 (2002).
\bibitem{Fujisawa}
T.\ Fujisawa {\it et al}., 
Nature (London) {\bf 419}, 278 (2002).
\bibitem{Ono}
K.\ Ono {\it et al}., Science {\bf 297}, 1313 (2002).
\bibitem{Sigurdur}
S.\ I.\ Erlingsson {\it et al}., %Yu.\ V.\ Nazarov, and V.\ I.\ Fal'ko,
Phys.\ Rev.\ B {\bf 64}, 195306 (2001).
\bibitem{Sigurdur2}
S.\ I.\ Erlingsson and Yu.\ V.\ Nazarov, cond-mat/0202237.
\bibitem{Ono2}
K.\ Ono {\it et al}., in preparation.
\bibitem{DNP}
B.\ E.\ Kane {\it et al}., Phys.\ Rev.\ B {\bf 46}, 7264 (1992);
K.\ R.\ Wald {\it et al}., Phys.\ Rev.\ Lett.\ {\bf 73}, 1011 (1994);
D.\ C.\ Dixon {\it et al}., Phys.\ Rev.\ B {\bf 56}, 4743 (1997).
\bibitem{Machida}
T.\ Machida {\it et al}., % S.\ Ishizuka, T.\ Yamazaki, S.\ Komiyama,
                          % K.\ Muraki and Y.\ Hirayama,
Phys.\ Rev.\ B {\bf 65}, 233304 (2002).
\bibitem{Kronmuller}
S.\ Kronm\"uller {\it et al}., Phys.\ Rev.\ Lett.\ {\bf 81}, 2526 (1998);
{\bf 82}, 4070 (1999).
\bibitem{Smet}
J.\ H.\ Smet {\it et al}., Phys.\ Rev.\ Lett.\ {\bf 86}, 2412 (2001).
\bibitem{com0}
In double quantum dots, it depends on the spin direction whether
the spin blockade takes place \cite{Ono}. If the spin of an incident
electron in the first dot is parallel to that of the electron trapped
in the second dot, the former electron cannot tunnel in the second dot
due to the Pauli exclusion principle. If the spins are anti-parallel,
the spin blockade does not happen; the electron tunnels in the second dot
and finally out to lead $R$ immediately. The total current is the sum of the
currents with and without spin flip.
We consider the former current only.
\bibitem{com1}
We assume that $k_{\rm B}T$ is larger than $E_{\rm Z}$.
Otherwise, the spin flip of electrons only from
$| \downarrow \rangle$ to $| \uparrow \rangle$ is
available by $H_{\rm hf}+H_{\rm el-ph}$,
which increases the components of larger $J$ and $-M$.
The other flip could be caused by $e.g$.\ higher-order
tunnel processes between the dot and leads \cite{Averin}.
\bibitem{Khaetskii2}
A.\ V.\ Khaetskii {\it et al}.,  %D.\ Loss, and L.\ Glazman,
Phys.\ Rev.\ Lett {\bf 88}, 186802 (2002).
\bibitem{Khaetskii}
A.\ V.\ Khaetskii and Yu.\ V.\ Nazarov,
Phys.\ Rev.\ B {\bf 61}, 12 639 (2000); {\bf 64}, 125316 (2001).
\bibitem{Averin}
D.\ V.\ Averin and Yu.\ V.\ Nazarov,
Phys.\ Rev.\ Lett.\ {\bf 65}, 2446 (1990);
M.\ Eto, Jpn.\ J.\ Appl.\ Phys.\ {\bf 40}, 1929 (2001).
\bibitem{Dicke}
R.\ H.\ Dicke, Phys.\ Rev.\ {\bf 93}, 99 (1954).
\bibitem{Brandes}
T.\ Brandes, in {\it Interacting Electrons in Nanostructures}, eds.\
R.\ Haug and H.\ Schoeller (Springer, Berlin Heidelbeg, 2001).
\bibitem{Brandes2}
T.\ Brandes {\it et al}., %J.\ Inoue and A.\ Shimizu,
Phys.\ Rev.\ Lett.\ {\bf 80}, 3952 (1998).
\end{references}
\end{document}